\documentclass[alpha-refs]{wiley}
\usepackage{framed}

\usepackage{siunitx}
\usepackage{todonotes}
\usepackage{hyperref}
\usepackage{wrapfig}
\usepackage{booktabs}
\usepackage{multirow}
\usepackage{wrapfig}

\newcommand{\ignore}[1]{}

\papertype{Research Article}

\title{AN INTERDISCIPLINARY EXPLORATION OF TRADE-OFFS BETWEEN ENERGY, PRIVACY AND ACCURACY ASPECTS OF DATA}

\author[1]{Pepijn de Reus}
\author[1]{Kyra Dresen}
\author[1]{Ana Oprescu}
\author[2]{Kristina Irion}
\author[3]{Ans Kolk}

\affil[1]{University of Amsterdam, Institute for Informatics, Faculty of Science}
\affil[2]{University of Amsterdam, Institute for Information Law, Amsterdam Law School}
\affil[3]{University of Amsterdam, Amsterdam Business School, Faculty of Economics and Business}

\corraddress{Ana Oprescu, University of Amsterdam, Amsterdam, The Netherlands}
\corremail{a.m.oprescu@uva.nl}

\fundinginfo{We acknowledge funding from our University's Interdisciplinary Research Theme Responsible Digital Transformations.}

\runningauthor{Pepijn de Reus, Kyra Dresen, Ana Oprescu, Kristina Irion \& Ans Kolk}

\begin{document}

\begin{frontmatter}
\maketitle

\begin{abstract}
The digital era has raised many societal challenges, including ICT's rising energy consumption and protecting privacy of personal data processing. This paper considers both aspects in relation to machine learning accuracy in an interdisciplinary exploration. We first present a method to measure the effects of privacy-enhancing techniques on data utility and energy consumption. The environmental-privacy-accuracy trade-offs are discovered through an experimental set-up. We subsequently take a storytelling approach to translate these technical findings to experts in non-ICT fields. We draft two examples for a governmental and auditing setting to contextualise our results. Ultimately, users face the task of optimising their data processing operations in a trade-off between energy, privacy, and accuracy considerations where the impact of their decisions is context-sensitive.

\ignore{After having created various user narratives, we worked with a focus group of legal and business practitioners as well as policymakers to discover their various valuations. These inputs can be used to present technical results in way more suitable for decision-making. With a general audience we also explored novel, more structured ways in which their stories might be told.}

\keywords{Digital economy, energy consumption of machine learning, data privacy, privacy-enhancing techniques, synthetic data, $k$-anonymity, storytelling}
\end{abstract}

\end{frontmatter}


\newpage
\section*{Introduction}
Besides many benefits, the digital ecosystem also brings societal challenges. Weighing the potential and the shortcomings of novel digital technologies informs policymakers and practitioners moving forward, and researchers can provide crucial insights to facilitate decision-making~\citep{ciulli2023international}. This paper aims to contribute by zooming in on two crucial implications of data-driven machine learning, i.e., the energy consumption and privacy. Both concerns have been raised in tandem prominently within the European Union (EU), which is the setting in which this study is placed. We undertake an interdisciplinary exploration of the trade-offs between the energy, privacy, and accuracy aspects of data processing using an experimental set-up (section 2). To facilitate decision-making we follow up this exploration with a storytelling approach (section 3), translating the technical findings to experts and decision-makers outside field of computer science. Our method, which extends earlier work in computer science that examines the change in energy consumption and accuracy of synthetic data and $k$-anonymity in machine learning tasks~\citep{oprescu2022energyk, de2023energy}, will be presented in section 2. First, however, we will indicate the broader context concerning the environmental and privacy aspects of the digital economy from the EU perspective.

A cornerstone in EU environmental policy is the European Green Deal and the "Fit for 55" package to reduce greenhouse gas (GHG) emissions by at least 55\% before 2030. Interestingly, the EU has connected its overall aim to remain within 1.5 degrees of global warming to digitalisation, by stating that “the twin challenge of a green and digital transformation has to go hand-in-hand”~\citep{european2020shaping}. Although this strategy seeks to harness digital technologies for achieving environmental goals, in this way also furthering the EU’s position in the digital realm, a concretisation is largely absent. Given that ICT’s carbon footprint is substantive, except if renewable energy is used, and will continue to rise in the future with the growth of the digital economy~\citep{freitag2021real}, optimising the energy consumption of data processing is imperative. This applies especially to standard data processing operations. Nevertheless, current research to reduce the energy consumption of ICT and data processing are mostly targeted at academic readers~\citep{verdecchia2023systematic}.

In addition to this increased focus on greening computing, data processing can trigger concerns over individuals’ privacy. In the EU, individuals’ personal data is protected by the fundamental rights to private life and to the protection of personal data. The General Data Protection Regulation (GDPR) provides for a corresponding legal framework for the processing of personal data~\citep{regulation2016regulation}. Using anonymous or anonymised data by contrast is no longer governed by the GDPR. Anonymisation techniques are considered privacy-enhancing technologies (PETs) that provide an important entry point for data operations that would otherwise be tightly regulated.

Combining these two societal concerns and EU policies raises the question of what privacy-enhancing algorithms add to the energy consumption of regular machine learning tasks. Moreover, there are perceived trade-offs between computing anonymised data and the accuracy of machine learning algorithms trained on this data. Ultimately, users face the task of optimising their data processing operations in a trade-off between privacy, accuracy, and environmental considerations. In the context of considering both the increased use of data and its climate impact, understanding which method would be preferred when using anonymisation algorithms ($k$-anonymity or synthetic data) in light of the energy consumed and the accuracy achieved is important. These are the aspects we will turn to in the next section.

\section{Key concepts and a method to assess trade-offs}
Several concepts from computer science form the background for this work, coherent to our approach we aim to make these comprehensible for a larger audience. This section provides an overview of anonymisation techniques that are data privacy-preserving, the energy consumption of code as a function of digital sustainability, and how the treatment of data impacts the accuracy of machine learning models. In a subsequent step, these concepts are related to each other to explore the trade-off between the environmental, privacy, and accuracy aspects of data.

\subsection{Data privacy}
Big data coins that the generation and processing of data are accelerating at scale~\citep{mayer2013big}. Training machine learning algorithms for instance requires large quantities of data. Data about individuals are called personal data and its use is in most countries regulated under personal data protection laws~\citep{greenleaf2023global}, a development heavily influenced by the standard-setting role of the EU’s GDPR (also called the ‘Brussels effect’, see \cite{bradford2020brussels}; \cite{coche2023unravelling}). The GDPR applies to the processing of personal data belonging to an identified or identifiable individual. When the GDPR applies, processing personal data has to be lawful and thus conforms to a number of legal requirements that aim to protect individuals’ rights and interests~\citep{oostveen2018golden}. Hence, the qualification as personal data or as non-personal data has important consequences for the application of the GDPR~\citep{finck2020they}. As a sub-set of privacy-enhancing technologies, anonymisation techniques are a means to remove personal data from datasets. After being treated with anonymisation techniques, the resulting dataset ideally becomes anonymised data and therefore non-personal data to which the GDPR no longer applies~\citep{finck2020they}. The advantage is evident because individuals’ data privacy is no longer an issue with anonymised data that can than be processed without further restrictions. What is more of an issue is whether the anonymisation of personal data is considered effective and irreversible~\citep{AEPD2023}. Out of various anonymisation techniques, $k$-anonymity and synthetic data are briefly outlined below.

\subsection*{$k$-anonymity}
\begin{wrapfigure}{R}{0.5\textwidth}
    \centering
    \includegraphics[width=0.48\textwidth]{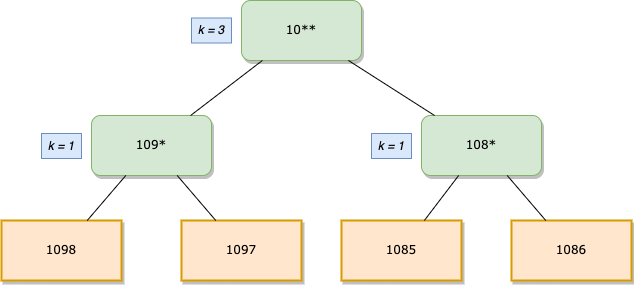}
    \caption{Anonymising of ZIP codes by suppression of the last digits. The more digits are suppressed, the more ZIP codes become indistinguishable from each other. The corresponding $k$-value is indicated in blue. Obtained from \cite{de2023energy}}
    \label{Suppression}
\end{wrapfigure}
$k$-anonymity is an algorithm that anonymizes the data by leaving out unique data features that can be traced back to a single person~\citep{samarati1998generalizing, sweeney2002k}. For any given value, called $k$, the algorithm will ensure that for this value at least $k$+1 individuals are indistinguishable from each other~\citep{sweeney2002k}. For example, a dataset with a $k$-anonymity of 5 ensures that for each data subject in the set, at least 5 other data subjects share the exact same properties. Because these 6 data subjects all have the same properties, they would thus be indistinguishable from each other. This process is exemplified in Figure~\ref{Suppression}.

The $k$-anonymity algorithm utilizes generalisation and suppression which are applied to hierarchies of attributes in the data. Once the data attributes are labeled using the data types above and a value for $k$ is chosen, the algorithm uses generalisation and suppression to alter the data~\citep{samarati2001protecting}. To come to this $k$-anonymity, all attributes in a structured dataset should be classiﬁed as one of four data types:
\begin{itemize}
    \item \textbf{Insensitive} attributes are deemed unimportant for privacy and will remain unaltered.
    \item \textbf{Sensitive} attributes are important for the subject (think of political views).
    \item \textbf{Identifying} attributes such as names are directly linked to a person and will be removed from the data.
    \item \textbf{Quasi-identifying} attributes refer to the data that could compromise privacy when linked with other attributes or datasets.
\end{itemize}

The algorithm leaves as much data as is while increasing the hierarchy of attributes in the dataset until the desired $k$-anonymity is obtained or until the entire attribute is suppressed. Yet, quasi-identifying attributes can still be unique trackers for this attribute. A known example is that a university has only one rector. If a university were to publish its salaries, staﬀ functions would be quasi-identifying. However, the function of a rector is a unique tracker as just one person fulﬁls this job. So even though quasi-identifying features in general are not identifying, they could be in case of a unique value. This forms a risk for re-identiﬁcation and, in turn, a compliance risk under the GDPR. Mindful that applying the $k$-anonymity algorithm is not by and in itself a guarantor for obtaining truly anonymised data, it really depends on the properties of the original dataset, the correct labeling of all attributes, setting an appropriate $k$-variable, and testing for re-identification.

\subsection*{Synthetic data}
An increasingly popular method to share data without comprising the data subjects' privacy is synthetic data~\citep{hernandez2022synthetic}. Synthetic data is a form of machine learning that learns the properties, co-variance, and distributions of a dataset and produces a representation of an original dataset. The synthetic dataset exhibits very similar statistical properties to the original dataset. Synthetic data hence mimics the properties of a dataset by creating artiﬁcial data that has a statistical resemblance to the original dataset without actually having overlap. After the synthetic data has been constructed, it should no longer contain personal data relating to individuals that are present in the original data. Because of this property, it is possible to publish and share synthetic data without compromising any individual’s privacy. Most importantly, the performance of machine learning on the synthetic dataset should be similar to that of the original dataset~\citep{abadi2016deep}. As synthetic data bypasses privacy and security protocols, it enables organisations to access data more rapidly~\citep{hittmeir2019utility}.

The first use cases of synthetic data are promising. For example in healthcare settings, where doctors and researchers often work with medical data from patients, synthetic data is increasingly used to share medical data~\citep{murtaza2023synthetic}. Or the Digital Twin project from the EU, where a digital copy is made to run new models and predictions~\citep{bauer2021digital}.

\subsection*{Digital sustainability}
Global heating is the result of GHG emissions, for example during energy generation of conventional energy options~\citep{schneider1989greenhouse}. One solution for this problem is to change to green energy generation. Another solution is to decrease energy consumption. Currently, we are operating on the basis of very rough estimations of what ICT's current share of GHG emissions is~\citep{mytton2022sources}. While conservative estimates attribute a 1.8\%–2.8\% share of global GHG emissions to ICT, other research suggests “that global emissions from ICT are as high as 2.1\%–3.9\%”~\citep{freitag2021real}. Yet, it is machine learning which is according to first studies~\citep{de2023growing} one of the biggest accelerators of ICT energy consumption and consequently ICT's claim of global GHG emissions. Where computing is powered by renewable energy this will not incur GHG emissions, however, also renewable energy is scarce and needs to be shared among all sectors of human activities.

Against this background, it is important to study ways of decreasing energy consumption of data processing operations. In the field of hardware, there is, according to Koomey's law~\citep{koomey2010implications}, an increase in the number of computations per Joule. However, this is not enough, because tasks are designed to need more computations to complete due to the confidence in the improvement of hardware~\citep{verdecchia2017estimating}. For this reason, we need to look at possibilities for decreasing energy consumption from a digital perspective. This is particularly important when we consider the two-century-old Jevons' paradox, which warns that increased efficiency leads to increased consumption. The increased consumption often exceeds the initial saving of efficiency, and has at least partially been reconfirmed~\citep{sorrell2009jevons}. In their research, \citet{fouquet2022twin}, analyse the speed of historical energy and communication technology transitions since 1850 which points to digitalisation outpacing the transition to renewable energy.

\subsection*{Model accuracy}
In the digital society progress in machine learning is appraised by looking at the accuracy of models performing
specific tasks. With machine learning, we measure the performance of models in various ways. For binary predictions where an answer can be either yes (1) or no (0), prediction accuracy is often used as metric~\citep{bratko1997machine}. Accuracy can be defined as the number of correct predictions divided by the number of predictions made. Hence an accuracy of 0.9 means that in 9 out of 10 cases, the model will predict the right target value. To achieve a high accuracy, machine learning methods optimize the parameters of their model by computing a loss function. This loss function is defined over all actions, where the goal of the model is to reduce the loss to the minimal possible~\citep{bishop2006pattern}. Using gradient updates, a model can reduce its loss by finding the optimal parameters. This is the learning part in machine learning: finding the optimal parameters to minimize loss and maximize the number of correct predictions.

\subsection{Technical trade-offs between environmental, privacy, and accuracy}
\label{sec:scenarios}
Data privacy, energy consumption, and model accuracy are important goalposts for data processing operations. But these concepts interact and affect each other. When users explore these different goalposts, they often perceive trade-offs that may prompt them to optimize their use case for just one of these values, casting aside the remaining two. Oftentimes users make decisions based on their perceptions about how these trade-offs play out. Comparing the values for each scenario, we examine the trade-offs and their intersection. We consider several three scenarios:
\begin{itemize}    
    \item \textbf{Scenario 0, optimise all dimensions}: \ignore{ideal so everyone wants this. But impossible for now and impossible for all applications, hence we have scenarios instead.} 
This is the ideal scenario, where all three goalposts are optimised simultaneously. The use of anonymisation techniques is oftentimes an escape valve from legislation such as the GDPR to obtain datasets that no longer contain personal data. In this scenario, users can choose an anonymisation technique that will produce the highest accuracy and consume the least energy. Although this field is still underexplored, recent work shows that $k$-anonymity can reduce the energy consumption of code as well as increase the accuracy in some cases~\citep{oprescu2022energyk}. However, it is not clear to what extent $k$-anonymity can satisfy the GDPR requirements. A recent paper indicates that with k-anon the identification risk is about 0.50\%~\citep{abubakar2022robustness}, meaning that for an average of 200 individuals in the data set, 1 person could be re-identified.\\

    \item \textbf{Scenario 1, optimise for accuracy}: If datasets that have been pre-processed with anonymisation techniques are perceived to negatively affect the accuracy of a given machine-learning model, then the user may disregard using PETs to preserve data privacy. This calculus could be corroborated by the energy consumption of applying anonymisation techniques to datasets. In this case, however, legislation on personal data protection such as the GDPR continues to apply which translates in compliance obligations and legal risks of non-compliance.\\

    \item \textbf{Scenario 2, optimise for energy}: When optimising for environmental considerations, the user would seek the least energy-consuming data processing techniques for the anonymisation of the dataset and subsequent model training. Using PETs should not be disregarded. It is also an option when the task at hand does not require the highest accuracy levels. A systemic review shows that the energy consumption of AI can often by reduced at least 50\%~\citep{verdecchia2023systematic}. In one specific dataset the energy consumption could be reduced by 77\% with the accuracy decreasing from 94.3\% to 93.2\%~\citep{brownlee2021exploring} which is negligible for most use cases.
\end{itemize}

\section{Experiment}
To delve into the trade-off, we set up an experiment to measure the impact of PETs on accuracy and the environment, with two leading research questions:
\begin{description}[noitemsep, topsep=0pt]
\item[\textbf{RQ1:}] Which privacy-enhancing technique yields the least loss of accuracy?
\item[\textbf{RQ2:}] Which privacy-enhancing technique is the least energy consuming?
\end{description}

In this section, we will first present the experimental setup, after which we present the results. Following these results, we link the findings to the presented trade-off.

\subsection{Experimental setup}
We reuse the setup of related work~\citep{de2023energy, oprescu2022energyk} to measure both the energy consumption and the impact on the accuracy for three machine learning techniques. We use two public datasets~\citep{BacheLichman2013}: the Student Performance set to predict a fail or pass of a student and the Census Income set to predict whether a person's income is above or below 50 thousand dollars. By varying three different datasets - original data, anonymised data (using $k$-anonymity) or synthetic data - we can compare the impact of each value on the trade-off. As we keep the other variables the same, the privacy enhancement is the control variable in this experiment. The setup and code are provided online at the public code repository GitHub\footnote{\url{https://github.com/PepijndeReus/Privacy-Enhancing-ML}}. The results of this experiment give insights into the loss of accuracy (\textbf{RQ1}). Over this entire process we measure the energy consumption by reading the Running Average Power Limiters (RAPL) counters (see details below), which helps to answer our research question for the energy consumption (\textbf{RQ2}).

\subsubsection{RAPL counters}
Computers with Intel processors from 2012 onward have energy sensors called RAPL in their Central Processing Units (CPU)~\citep{desrochers2016validation, hahnel2012measuring}. These RAPL sensors measure the energy consumption of the CPU and memory of the Intel chip. An advantage of RAPL is that the user can select which parts of the chip should be measured~\citep{garcia2019estimation, khan2018rapl}, for example solely the CPU or the memory energy consumption. Another option to measure the energy consumption would be to use a hardware-based method, also known as the power plug method. This is an external device between the socket and the used machine. Such a device is very accurate, but can measure only the entire energy input to the device and hence not divide between CPU or memory consumption. Apart from that, the devices are expensive and costly to use regarding set-up time. Hence, there are various methods to measure the energy consumption of code.

\subsection{Statistical significance}
We can show the significance of the impact on energy consumption due to the addition of the code smell, proving that the samples from the original and from the smelly measurements are from different distributions for a specific problem written in a specific language. Additionally, we wish to know the direction of the impact. 

The aim of a statistical test is to reject a null hypothesis in favor of an alternative hypothesis. By rejecting the null hypothesis we can claim that the alternative hypothesis is significantly true. For this purpose, we derive some statistics from our sample distribution called the p-value. This p-value is the probability of obtaining a result given that the null hypothesis is true. If this p-value is smaller than a predetermined significance level ($\alpha$-value), we may reject the null hypothesis. The significance level indicates the probability of rejecting the null hypothesis when it is true. Typically, the significance level is set at 0.05.

\subsubsection{The Mann-Whitney U test}
The Mann-Whitney U test~\citep{mann1947test} looks at the probability that a sample from one distribution is greater than a sample from another distribution. If this probability is at 50\%, the two samples belong to the same distribution. There are two Mann-Whitney U tests: the one-sided and two-sided tests. While the two-sided test has an alternative hypothesis claiming that the two samples are from different distributions, the one-sided test has an alternative hypothesis claiming that the first sample is from a greater distribution~\citep{nachar2008mann}.

To find the direction of the difference between the distributions, we must use the one-sided test twice. We use the greater one-sided Mann-Whitney U test to determine the probability that a sample from the first distribution is larger than a sample from the second distribution. The lesser one-sided Mann-Whitney U test determines the probability that a sample from the first distribution is smaller than a sample from the second distribution. If this p-value of the \textit{greater} one-sided test is smaller than our significance level, we can say that the first distribution is significantly \textit{smaller} than the second distribution, and vice versa for the lesser one-sided test.

\subsection{Technical results}
In Table~\ref{ResultsEnergyCon} (see next page), we present the results of the energy consumption for the Census Income dataset on the left and the Student Performance set on the right. The benchmark is presented on the first row, after which the results are shown for the models trained on either anonymised or synthetic data. We present the deviation in percentages to the benchmark for these methods. We use several exemplars of machine learning techniques: $k$-nearest neighbours (knn) -- a non-parametric approach, logistic regression (LogReg) -- a parametric approach, and Neural Network (NN) -- a simple, feedforward neural network. Complementary with the energy consumption, we provide the results of the Mann-Whitney U test in Table~\ref{Paper:MannWhit}. Finally, the idle energy consumption of the machine is measured. For each second of inactivity, 7.512 Joules are consumed by the machine, to be subtracted from the results. However, since we look at the trends this idle energy consumption is deemed negligible.

Table~\ref{PaperAccModels} shows the accuracy score of the machine learning techniques for each dataset and the privacy-enhancing technique. The benchmark reflects the accuracy scores for the unaltered or original data. To reflect the accuracies as exactly as possible we present the accuracy scores instead of the deviation to the benchmark. As $k$-anonymity is obtained via generalisation and suppression, some data has been suppressed entirely. Table~\ref{PaperSuppressedData} shows the percentages of suppressed data for each dataset and respective $k$-value.

\begin{table*}[h!]
\centering
\caption{The average run time and energy consumption for the Census Income and Student Performance datasets. On the rows, we define the input data and on the columns, we define each method, devised into run time and energy consumption. The benchmark resembles the unaltered data as obtained online, the percentages below show the deviation for each method to this benchmark.}
\label{ResultsEnergyCon}
\resizebox{13cm}{!}{
\begin{tabular}{lcccccclcccccc}
\toprule
\multicolumn{7}{c}{\textbf{Census Income dataset}} & \multicolumn{7}{c}{\textbf{Student Performance set}} \\
\cmidrule(r){1-7} \cmidrule(r){8-14}

    & \multicolumn{2}{c}{knn} & \multicolumn{2}{c}{LogReg} & \multicolumn{2}{c}{NN} &   & \multicolumn{2}{c}{knn} & \multicolumn{2}{c}{LogReg} & \multicolumn{2}{c}{NN} \\
\cmidrule(l{3pt}r{3pt}){2-3} \cmidrule(l{3pt}r{5pt}){4-5} \cmidrule(l{5pt}r{2pt}){6-7} \cmidrule(l{3pt}r{5pt}){9-10} \cmidrule(l{5pt}r{2pt}){11-12} \cmidrule(l{5pt}r{2pt}){13-14}
Benchmark & 8.19s & 329.33J & 1.41s & 73.70J & 9.85s & 157.59J & Benchmark & 0.04s & 2.16J & 0.06s & 2.95J & 3.40s & 61.20J \\
$k$=3 & -67\% & -67\% & -45\% & -46\% & -44\% & -42\% & $k$=3 & -25\% & -37\% & -66\% & -77\% & -26\% & -24\% \\
$k$=10 & -69\% & -68\% & -61\% & -62\% & -50\% & -47\% & $k$=10 & -75\% & -79\% & -66\% & -78\% & -28\% & -27\% \\
$k$=27 & -73\% & -73\% & -78\% & -78\% & -54\% & -51\% & $k$=27 & -75\% & -80\% & -66\% & -79\% & -29\% & -26\% \\
Synthetic data & -5\% & -4\% & -9\% & -10\% & -1\% & 0\% & Synthetic data & 0\% & +2\% & -33\% & +3\% & -1\% & -2\% \\
\bottomrule
\end{tabular}}
\end{table*}
\vspace{-0.5cm}
\begin{table*}[h!]
\centering
\caption{The results of the Mann-Whitney U test applied to the energy consumption presented in Table~\ref{ResultsEnergyCon}. The presented p-value shows the probability of the method on the row being greater than the alternative in the column.}
\label{Paper:MannWhit}
\resizebox{11cm}{!}{
\begin{tabular}{lcccclcccc}
\toprule
\multicolumn{5}{c}{\textbf{Census Income dataset}} & \multicolumn{4}{c}{\textbf{Student Performance set}} \\
\cmidrule(r){1-5} \cmidrule(r){6-10}
    & $k$=3 & $k$=10 & $k$=27 & Synthetic data &    & $k$=3 & $k$=10 & $k$=27 & Synthetic data \\
$k$=3 & x & p \textless 0.01 & p \textless 0.01 & p \textgreater 0.99 & $k$=3 & x & p \textless 0.01 & p \textless 0.01 & p \textgreater 0.99 \\
$k$=10 & p \textgreater 0.99 & x & p \textless 0.01 & p \textgreater 0.99 & $k$=10 & p \textgreater 0.99 & x & p \textless 0.01 & p \textgreater 0.99 \\
$k$=27 & p \textgreater 0.99 & p \textgreater 0.99 & x & p \textgreater 0.99 & $k$=27 & p \textgreater 0.99 & p \textgreater 0.99 & x & p \textgreater 0.99 \\
Synthetic & p \textless 0.01 & p \textless 0.01 & p \textless 0.01 & x & Synthetic & p \textless 0.01 & p \textless 0.01 & p \textless 0.01 & x \\
\bottomrule
\end{tabular}}
\end{table*}

\begin{table}[h!]
\parbox{.55\linewidth}{
\caption{The weighted accuracy over 10 measurements for each dataset, machine learning technique, and privacy-enhancing technique.}
\label{PaperAccModels}
\resizebox{7cm}{!}{
\begin{tabular}{lcccccc}
\toprule
    & \multicolumn{3}{c}{\textbf{Census Income dataset}} & \multicolumn{3}{c}{\textbf{Student Performance set}} \\
\cmidrule(r){2-4} \cmidrule(r){5-7}
    & knn & LogReg & NN & knn & LogReg & NN \\
Benchmark & 0.820 & 0.846 & 0.846 & 0.700 & 0.719 & 0.706 \\
$k$=3 & 0.828 & 0.848 & 0.847 & 0.872 & 0.851 & 0.861 \\
$k$=10 & 0.832 & 0.842 & 0.842 & 0.826 & 0.826 & 0.826 \\
$k$=27 & 0.828 & 0.837 & 0.837 & 0.831 & 0.853 & 0.853 \\
Synthetic data & 0.802 & 0.828 & 0.828 & 0.728 & 0.756 & 0.755 \\
\bottomrule
\end{tabular}}
}
\hfill
\parbox{.35\linewidth}{
\caption{The percentage of suppressed data for the Census Income set and Student Performance with respect to the obtained $k$-anonymity. The percentage is defined as suppressed cells divided by the total amount of cells.}
\label{PaperSuppressedData}
\resizebox{5cm}{!}{
\begin{tabular}{lcc}
\toprule
\multicolumn{3}{c}{\textbf{Suppressed data}} \\
\cmidrule(r){2-3}
& Census Income set & Student Performance set \\
$k$=3 & 19\% & 74\% \\
$k$=10 & 28\% & 80\% \\
$k$=27 & 32\% & 84\% \\
\bottomrule
\end{tabular}}
}
\end{table}

\subsection{Interpreting the results to the trade-off}
The results of this experiment show that models trained on $k$-anonymised data consume less energy than models trained on the original data, with a similar accuracy to the unaltered data. Models trained on synthetic data have a similar energy consumption and a similar to lower accuracy compared to models trained on unaltered data.

These findings are interesting in the light of the presented trade-off. Apparently, choosing data privacy using either $k$-anonymity or synthetic data does not lead to a substantial reduction in accuracy. In fact, the accuracy might even improve. Similarly, machine learning models trained using PETs have a similar to lower energy consumption compared to models trained on unaltered data. As these findings form an interesting argument against the perceived trade-off, we wish to translate these scientific results into findings suitable for non-computer science experts. And so follows our final research question: how can we translate these technical results to non-computer science experts?

\section{Translating the findings to non-computer science experts}
Data can be hard to value, especially when comparing the value of data across different fields of research. For example, lawyers may prioritise compliance with regulations on data privacy over accuracy whereas data scientists might strive for the highest accuracy, but should not ignore legal compliance. To link these different valuations of the presented trade-off, we explore the economic contextualisation of our presented technical results using ~\cite{slotin2018we} recommendation 
that when it comes to measuring the value of data often an impact-based valuation works best. The impact-based method demonstrates the relationship between data and outcomes in people's lives to determine how to value data. A typical impact-based approach is storytelling.

\subsection{Introduction to storytelling}
Often a story is defined as something with a beginning, a middle, and end~\citep{moezzi2017using}. In \textit{storytelling} the emphasis is on the elicitation and construction of stories or narratives in situ. An important aspect is why the story is told in a particular context. Stories are crafted rather than pre-existing artefacts, and crafting depends on various contexts such as audience, purpose, location, etc.~\citep{moezzi2017using}. Storytelling in science is important to translate scientific findings to policymakers, journalists, and the general public~\citep{schimel2012writing}. In his book on Writing Science, ~\citet[p~11]{schimel2012writing} states: "The data are supporting actors in the story you tell. The lead actors are the questions and the larger issues you are addressing. The story grows from the data, but the data are not the story." This is why storytelling is a method that has the potential to explain the trade-offs between digital sustainability, data privacy, and model accuracy to a non-technical audience as they present themselves after our experiment. 

\subsection{Stories translating the experiment}

\ignore{Now suppose we have Bob, a healthcare data scientist working for a research hospital. Bob too works with sensitive datasets, comparable to the Census Income and Student Performance sets, for which he has to perform binary classification. However, in this case, the prediction task is high risk, namely whether patients should be considered for a research trial procedure. Bob, who has a mathematics background, could model this decision-making problem as a Pareto-optimality problem~\citep{kuosmanen2005measuring, huppes2005framework}, and give some (equal) weights to the three dimensions of interest to come to a solution. In this case, however, he has a hard time quantifying how many more times the privacy of a patient is more important than the accuracy of a treatment of a different patient. Consequently, Bob hesitates between synthetic data and $k$=27-anonymity, although for either of them, he would choose logistic regression as ML technique, since it delivers equally excellent accuracy as the Neural Network technique, but at a lower energy cost.}

The first story offers a narrative for optimising accuracy (Section~\ref{sec:scenarios}, Scenario 1). Suppose we have Alice, a data scientist working for a municipality. Alice is developing an algorithm to detect illegal holiday  rentals\footnote{\url{https://algoritmeregister.amsterdam.nl/ai-system/illegal-holiday-rental-housing-risk/109}}. In the current workflow, when a report is filed with the municipality that there may be a case of illegal renting at an address, the Surveillance \& Enforcement employees could start an investigation. To help this team, Alice's algorithm should assist them in prioritising the reports so that their team's enforcement capacity can be used efficiently. Thus, Alice works with sensitive datasets, comparable to the datasets used in our experiment, for which she has to initially perform binary classification (whether it is indeed a case of illegal holiday rental). Basically, the dataset is anonymised before the model is trained, and then the model is used to indicate how likely it is that a report of illegal rental is correct. So, in itself, the model does not detect illegal rental, but gives advice to the human expert. Both k-anonymity~\citep{de2023real} and synthetic data\footnote{\url{https://www.kaggle.com/datasets/ealaxi/paysim1}} have been used as anonymisation techniques in relation to fraud detection applications. Although the decision-making on the illegal rental case is not automated, and the municipality employee can still choose to ignore the output of the algorithm, the prediction task is high risk, as a visit from the authorities to one's household can be seen as very intrusive. In this story, both privacy and accuracy are important. Alice, who has a mathematics background, could model this decision-making problem as a Pareto-optimality problem~\citep{kuosmanen2005measuring, huppes2005framework}, and give some (equal) weights to the three dimensions of interest to come to a solution. She has a hard time however quantifying how many more times the privacy of a citizen is more important than the accuracy of an indication of fraud. Consequently, Alice hesitates between synthetic data and $k$=27-anonymity. For either of them, she would choose Logistic Regression as machine learning technique, since it delivers equally excellent accuracy as the Neural Network, but at a lower energy cost.

The second story offers a narrative about optimising energy (Section~\ref{sec:scenarios}, Scenario 2). Suppose we have Bob, working as a digital enabler for an auditing company. Bob too has to perform binary classification on sensitive datasets, comparable to the datasets used in our experiment. Though the datasets are sensitive, the prediction task is low risk, in the sense that its accuracy is not crucial. Bob needs to decide what PET to use and what machine learning model to pick. Bob wants to optimise over energy and make an informed decision. Next, Bob is informed by our technical findings~\citep{de2023energy}, and he decides that he would like to use synthetic data to enhance the privacy of the subjects in the sensitive datasets. Moreover, he wishes to minimize his climate impact by picking the lowest energy consumption. Consequently, Bob picks synthetic data and Logistic Regression to predict the binary labels for his dataset.

\subsection*{Limitations}
We conducted the technical experiments on two structured datasets. This is a limitation in two ways: 1) using more datasets would allow to generalise findings; 2) structured data is often a utopia in data science and anonymisation methods cannot (yet) be applied to unstructured data. Moreoever, several sectors such as healthcare have special regulations and protocols with respect to personal privacy which have to be observed. Our study does not cover these regulated sectors. Moreover, $k$-anonymity is no guarantee that the data has been truly anonymised. Anonymisation is highly context-specific and needs to be controlled for on a case-by-case basis. Finally, our work is aimed at the GDPR and therefore a Western perspective. We acknowledge that a matured version of our work should cover universal aspects and especially the trade-offs in the Global Souths~\citep{milan2019big}.

\section{Conclusion} 
With the EU backdrop of equal concern for protecting data privacy (GDPR) and reducing climate impact (Green Deal), it is becoming increasingly important to understand how society perceives the trade-off between these two dimensions and triangulate it with model accuracy. 
\ignore{We innovate by structuring storytelling with the support of Agile methodology. The stakeholder pre-interviews have indeed fostered the dialogue between business, governance, and software engineering research required by sustainable prosperity in the age of a Data Economy regarding the choices of privacy and climate impact, and how Government-to-Government and Business-to-Government will affect their scope.
}
In this study, we make a first effort to provide insight into the trade-off between privacy and energy consumption in light of machine learning. We explore the question of what privacy-enhancing algorithms, such as synthetic data and $k$-anonymity, add to the energy consumption of regular machine-learning tasks. Using an experiment to answer this technical question, we then form an economic contextualisation through story-telling of our technical results. With these stories we aim to make technical results accessible to non-ICT experts and decision-makers, and as such aim to understand how society could perceive the trade-off between privacy and climate impact in the digital economy.
We find that, in some contexts, such as the advertisement industry, the answers are clear-cut and simple. However, we also find that in some sensitive application domains, such as personal finance and healthcare, the legal compliance and ethical aspects require a different approach than the trade-off approach. To obtain answers in these contexts, we need to re-assess the way we typically evaluate the value of data. We provide suggestions for future work to close the gap between scientific results and policymakers.
\ignore{ Below the text that I edited to the piece that is now above here:
We find that in some contexts, the answers are clear-cut and simple. However, we also find that in some extremely sensitive application domains, such as healthcare, to obtain these answers we need to re-assess the way we typically evaluate the value of data.}

\subsection*{Future work}
Policymakers from non-ICT backgrounds should be able to make informed decisions on the digital economy. To discover the scenarios relevant to them, we plan to extend our research to devise novel story-developing methods by questioning experts from various fields and stakeholders from various value chains. Using these methods, research should be aimed towards storytelling as a means to share scientific insights. Apart from that, details on the energy consumption of the digital economy are often unknown. In parallel, more research efforts should be focused on measuring the energy consumption of various AI techniques and data operations, on the impact of considering different data utility metrics, and on refining the granularity at which we measure anonymity and privacy (using tools such Anonymeter\footnote{\url{https://github.com/statice/anonymeter}}. By expanding to more datasets, preferably from daily-use scenarios, we discover how the various trade-offs interact. Combining technical research with the right method to inform policymakers leads to informed decisions to reduce the impact of the digital economy on the environment.


\section*{Acknowledgements}
We would like to express our gratitude towards Bart Krull and Sustainalab\footnote{\url{https://sustainalab.nl}} for connecting us to the Climate Tech Cities community. We would also like to thank dr. Christine Erb from RDT for her continuous support of our ideas.



\section*{Conflict of Interest}
Kristina Irion is a co-author but also one of the organisers.







\section*{Data Availability Statement}
The data that support the findings of this study are openly available in UCI Machine Learning Repository at \url{http://www.archive.ics.uci.edu/}. 

\printendnotes

\bibliography{literature}

\end{document}